\documentclass[useAMS,usenatbib]{mnras}

\usepackage{natbib}
\usepackage{mathtools}
\usepackage{amsmath}
\usepackage{amssymb}

\title[M31 globular cluster ages]{Measuring M31 globular cluster ages and metallicities using both photometry and spectroscopy}

\author[Usher et al.]{Christopher~Usher,$^{1}$\thanks{email: chris.usher@astro.su.se}
Nelson~Caldwell,$^{2}$ Ivan Cabrera-Ziri,$^{3}$\\
$^{1}$The Oskar Klein Centre, Department of Astronomy, Stockholm University, AlbaNova, SE-106 91 Stockholm, Sweden \\
$^2$Harvard-Smithsonian Center for Astrophysics, 60 Garden Street, Cambridge, MA 02138, USA\\
$^3$Astronomisches Rechen-Institut, Zentrum f\"ur Astronomie der Universit\"at Heidelberg, M\"onchhofstra\ss e 12-14, D-69120 Heidelberg, Germany\\
}

\begin{document}
\maketitle

\begin{abstract}
The ages and metallicities of globular clusters play an important role not just in testing models for their formation and evolution but in understanding the assembly history for their host galaxies.
Here we use a combination of imaging and spectroscopy to measure the ages and metallicities of globular clusters in M31, the closest massive galaxy to our own.
We use the strength of the near-infrared calcium triplet spectral feature to provide a relatively age insensitive prior on the metallicity when fitting stellar population models to the observed photometry.
While the age-extinction degeneracy is an issue for globular clusters projected onto the disc of M31, we find generally old ages for globular clusters in the halo of M31 and in its satellite galaxy NGC 205 in line with previous studies.
We measure ages for a number of outer halo globular clusters for the first time, finding that globular clusters associated with halo substructure extend to younger ages and higher metallicities than those associated with the smooth halo.
This is in line with the expectation that the smooth halo was accreted earlier than the substructured halo.
\end{abstract}

\begin{keywords}
globular clusters: general, galaxies: star clusters: general, galaxies: stellar content, galaxies: evolution 
\end{keywords}

\section{Introduction}
 \label{sec:intro}

Globular clusters (GCs) are a powerful tool to study galaxy formation (see reviews by \citealt{2006ARA&A..44..193B} and \citealt{2018RSPSA.47470616F}) as the properties of GCs observed today reflect both the conditions of their formation but also the physics of GC survival.
Found in virtually all galaxies with stellar masses above $10^{9}$ M$_{\odot}$ (and most galaxies more massive than $10^{7}$ M$_{\odot}$, \citealt{2013ApJ...772...82H, 2022ApJ...926..162E}), they are significantly brighter than individual stars, allowing them to be studied at much greater distances.
The oldest GCs serve as fossils of the earliest stages of galaxy formation while the continuous formation of GCs until today \citep[e.g.][]{2018MNRAS.473.2688M, 2019MNRAS.489L..80B} allows them to trace the full history of galaxy formation.
GC ages provide both an important test of theories of how GCs formed and evolved but also can provide important constraints on how galaxies form and assemble.

In this paper we use the term GC for any star cluster older than $\sim 1$ Gyr and young massive cluster (YMC) for any star cluster younger than this age.
This division corresponds to the transition for a massive cluster from being dominated by formation physics to disruption physics and in stellar evolution terms to when the red giant branch starts to make a significant contribution to the luminosity of the cluster.

Two broad classes of models for GCs have been proposed.
Motivated by the old ages of Milky Way (MW) GCs \citep[e.g.][]{1983ApJ...264..206J, 2013ApJ...775..134V, 2021JCAP...08..017V} and the lack of easily observable star clusters forming today in the MW massive enough to survive a Hubble time, the first class of models \citep[e.g.][]{1984ApJ...277..470P, 2015ApJ...808L..35T} invoke special conditions in the early Universe.
In these models GCs form during or even before the epoch of reionisation ($z \gtrsim 6$, lookback times $\gtrsim 13$ Gyr), often in their own dark matter halos.
In the second class of models \citep[e.g.][]{1997ApJ...480..235E, 2010ApJ...712L.184E, 2015MNRAS.454.1658K}, motivated by observations of star clusters of GC mass or greater forming in starburst galaxies today \citep[e.g.][]{1995AJ....109..960W, 1996AJ....112..416H, 2020MNRAS.499.3267A}, GC formation is the natural outcome of intense star formation.
In the first class of models, all GCs in all galaxies should have formed in the first Gyr of the Universe; in the second, the distribution of GC ages should be broader and should vary with galaxy assembly history.
Quantitative models of GC formation \citep[e.g.][]{2014ApJ...796...10L, 2018MNRAS.475.4309P, 2021MNRAS.505.5815V, 2023MNRAS.521..124R} make predictions for the age distribution of GCs that can be compared to observations.

GC are also powerful probes of the formation and assembly history of galaxies.
Since a large star formation rate density is required to form a massive, compact star cluster, the ages of GCs trace periods of intense star formation.
If the age and metallicity of a GC are known, the galaxy mass-metallicity relation \citep[see review by][]{2019A&ARv..27....3M} can be inverted to give the likely mass of the galaxy the GC formed in \citep{2019MNRAS.486.3134K}.
With multiple GCs, the mass assembly history of a galaxy can be constrained.
The presence of GCs with a significant range of metallicities at the same age implies those GCs formed in different mass galaxies.
The presence of at least two branches of the MW GC age-metallicity relationship was used to argue that the older branch was the in situ population and the younger branch was the accreted population \citep{2010MNRAS.404.1203F, 2013MNRAS.436..122L}.
This association was confirmed by the younger branch GCs having the same orbits as accreted satellite galaxies \citep[e.g.][]{2019A&A...630L...4M}
By folding in constraints on the orbits of GCs, the merger tree of a galaxy can be reconstructed in detail as was done for the MW by \citet{2020MNRAS.498.2472K} since earlier mergers and more massive mergers should deliver their GCs to smaller galactocentric radii \citep{2020MNRAS.499.4863P}.

Beyond the Local Group, GCs cannot be resolved into their constituent stars and have to be studied using their integrated light. 
Despite their importance, measuring reliable GC ages from their integrated light remains a challenge.
Optical photometry suffers from strong age-metallicity and age-extinction degeneracies \citep[e.g.][]{1994ApJS...95..107W, 2004MNRAS.347..196A, 2014A&A...569A...4D} in that old ages and higher metallicities both make a stellar population redder as can the effects of interstellar extinction. 
While UV or NIR photometry can be used to break this degeneracy, obtaining deep enough photometry for a significant sample of GCs can be observationally expensive.
Spectroscopy can also provide ages but the quality of spectra required is likewise observationally expensive for significant samples.
While spectroscopy does not suffer from the age-extinction degeneracy, it can suffer from a degeneracy between the morphology of the horizontal branch and age; see \citet{cabrera-ziri22} for a detailed discussion.

\citet{2019MNRAS.490..491U} used optical photometry and spectroscopy of the near infrared calcium triplet (CaT) to derive ages and metallicities for GCs in three massive early-type galaxies from the SLUGGS survey \citep{2014ApJ...796...52B} and the MW.
\citet{2019MNRAS.482.1275U} showed that the strength of the CaT is a reliable and age-independent measure of metallicity for stellar populations older than a couple Gyr.
\citet{2019MNRAS.490..491U} used these metallicities as priors when fitting the photometry with stellar population models to measure ages.
In line with previous work \citep[e.g.][]{2015MNRAS.446..369U, 2016ApJ...829L...5P} that found that the GC colour-metallicity or colour-colour relationship varies between galaxies, \citet{2019MNRAS.490..491U} found these four galaxies show widely different distributions of GCs in age-metallicity space, suggesting different assembly histories and disfavouring models where the majority of GCs form in the early Universe.

As the nearest (780 kpc, \citealt{2012ApJ...758...11C}) massive (stellar mass $\sim 10^{11}$ M$_{\odot}$, \citealt{2012A&A...546A...4T}) galaxy to the Milky Way, M31 provides an important test of techniques for studying GCs using their integrated light.
M31 is close enough that its GCs can be observed in ways that are impractical for more distant galaxies - its GCs can be resolved into their constituent stars using space based observations and high signal-to-noise ratio, high resolution integrated spectroscopy can be obtained in a reasonable amount of time - yet it is distant enough that is relatively straight forward to obtain integrated photometry and spectroscopy of its GCs.

There is a long history of studying the metallicities and ages of GCs in M31 using their integrated light 
\citep[e.g.][]{1969ApJS...19..145V, 1984ApJ...287..586B, 2005A&A...434..909P, 2005AJ....129.1412B, 2009A&A...508.1285G, 2011AJ....141...61C, 2022MNRAS.512.4819S}.
The superior spatial resolution of the Hubble Space Telescope (HST) allows the star clusters of M31 to be resolved into their individual stars.
In most cases the resulting colour-magnitude diagrams only reach the upper red giant branch \citep[e.g.][]{1997AJ....114.1488H, 2005AJ....129.2670R, 2009A&A...507.1375P} but in the case of B379-G312, the photometry reaches below the main sequence turnoff, allowing the age to be measured directly \citep{2004ApJ...613L.125B, 2010PASP..122.1164M}.
With shallower photometry, lower limits on the age can be placed and the presence of a blue horizontal branch can be used as evidence for an old stellar population \citep[e.g.][]{2011A&A...531A.155P}.
While most M31 GCs seem to be old ($\gtrsim 10$ Gyr) and M31 hosts a population of YMCs ($\lesssim 1$ Gyr e.g. \citealt{1979AJ.....84..744H, 1988ApJ...333..594E, 2009AJ....137...94C, 2016ApJ...827...33J}), the existence of a population of younger GCs  (between 1 and 10 Gyr) has been debated \citep[e.g.][]{2004ApJ...614..158B, 2005AJ....129.1412B, 2011AJ....141...61C, 2011A&A...531A.155P}.

In this paper we extend the analysis of \citet{2019MNRAS.490..491U} to M31 by using literature spectra described in Section~\ref{sec:data} to measure the metallicity of 290 GCs using the strength of the CaT spectral feature (Section~\ref{sec:CaT}).
In Section~\ref{sec:age} we use these metallicities as a prior when deriving their ages from their optical photometry before summarising our work in \ref{sec:discussion}.

\section{Literature Data}
\label{sec:data}

\subsection{Spectroscopy}
We used spectra of the CaT region from \citet{2009AJ....137...94C} and from \citet{2016MNRAS.456..831S}, \citet{2021MNRAS.502.5745S} and \citet{2022MNRAS.512.4819S}.
\citet{2009AJ....137...94C} observed over 500 star clusters in M31 using the Hectospec mulitfibre spectrograph \citep{2005PASP..117.1411F}. 
The Hectospec spectra cover 3700 to 9200 \AA{} at a resolution of $\sim 5$ \AA{} and a dispersion of 1.2 \AA .
We restricted ourselves to the subsample of 316 star clusters determined by \citet{2011AJ....141...61C} to be old.
These Hectospec spectra all lie within 20 kpc in projection of the centre of M31.

\citet{2016MNRAS.456..831S} observed 27 GCs at a range of galactocentric distances using the Dual Imaging Spectrograph (DIS) on the 3.5 m telescope at Apache Point Observatory and 5 GCs using the Sparsepak fibre unit \citep{2004PASP..116..565B, 2005ApJS..156..311B} and the Bench Spectrograph \citep{2008SPIE.7014E..0HB, 2010SPIE.7735E..7DK} on the WIYN 3.5 m telescope.
\citet{2021MNRAS.502.5745S} observed a single GC (G001-MII) with DIS and the Apache Point Observatory 3.5 m telescope and \citet{2022MNRAS.512.4819S} observed 30 GCs in the outer halo of M31 with same instrument and telescope.
The DIS spectra cover 8000 to 9100 \AA{} at resolution of $R \sim 4000$; the WIYN spectra cover 8300 to 8800 \AA{} at resolution of $R \sim 9000$.
We do not include the spectrum of B457-G097 from \citet{2016MNRAS.456..831S} in our analysis since \citet{2016MNRAS.456..831S} suggest that they observed a different object given the substantial differences in radial velocity ($\sim 300$ km s$^{-1}$) and metallicity ($\sim 0.5$ dex) between their measurements and literature values; we do use the spectrum of B457-G097 from \citet{2011AJ....141...61C}.
Nor do we include the spectrum of dTZZ-05 as suggested by \citet{2022MNRAS.512.4819S} since the measured radial velocity is more similar to a MW star than a M31 GC.

\subsection{Imaging}
We preferred the Sloan Digial Sky Survey (SDSS, \citealt{2000AJ....120.1579Y}) $ugriz$ photometry of \citet{2010MNRAS.402..803P} when available.
For more recently discovered GCs in the SDSS footprint, we used the SkyServer\footnote{http://skyserver.sdss.org/dr17} to download SDSS DR17 \citep{2022ApJS..259...35A} \textsc{model} photometry for each GC. 
We applied the 0.04 mag correction to bring the $u$-band onto the AB system.
For halo GCs not covered by SDSS, we used the PAndAS Canada France Hawaii Telescope MegaCam $gi$ photometry from \citet{2014MNRAS.442.2165H}.

\section{Calcium triplet based metallicities}
\label{sec:CaT}
We measure the metallicities of the M31 GCs using the strength of the calcium triplet (CaT) rather than relying on the \citet{2011AJ....141...61C} Lick index based metallicities to both maintain commonality with \citet{2019MNRAS.490..491U} and since the strength of the CaT shows little dependence on age \citep[e.g.][]{2003MNRAS.340.1317V, 2019MNRAS.482.1275U}.
We measure the strength of the CaT using the technique of \citet{2010AJ....139.1566F} and \citet{2012MNRAS.426.1475U, 2019MNRAS.482.1275U}.
We use the same Python code as in \citet{2019MNRAS.482.1275U} to fit the observed spectra with a linear combination of stellar templates (we use the same templates as \citealt{2010AJ....139.1566F, 2012MNRAS.426.1475U, 2019MNRAS.482.1275U}), normalise the continuum (using the same parameters as in \citealt{2019MNRAS.482.1275U}) and measure the combined CaT strength of all three lines from the fitted templates using the same index definition (that of \citealt{1988AJ.....96...92A}) as in our previous work.
We refer the interested reader to \citet{2019MNRAS.482.1275U} for details of the measurement process and a detailed discussion of potential systematics.
To avoid systematics due to poor sky subtraction or low signal-to-noise ratios (S/N), we only measure spectra with S/N greater than 20 \AA $^{-1}$.
For the Sakari et al. spectra we also excluded a handful of spectra with higher S/N that showed strong sky subtraction residuals.

\begin{figure}
    \centering
    \includegraphics[width=216pt]{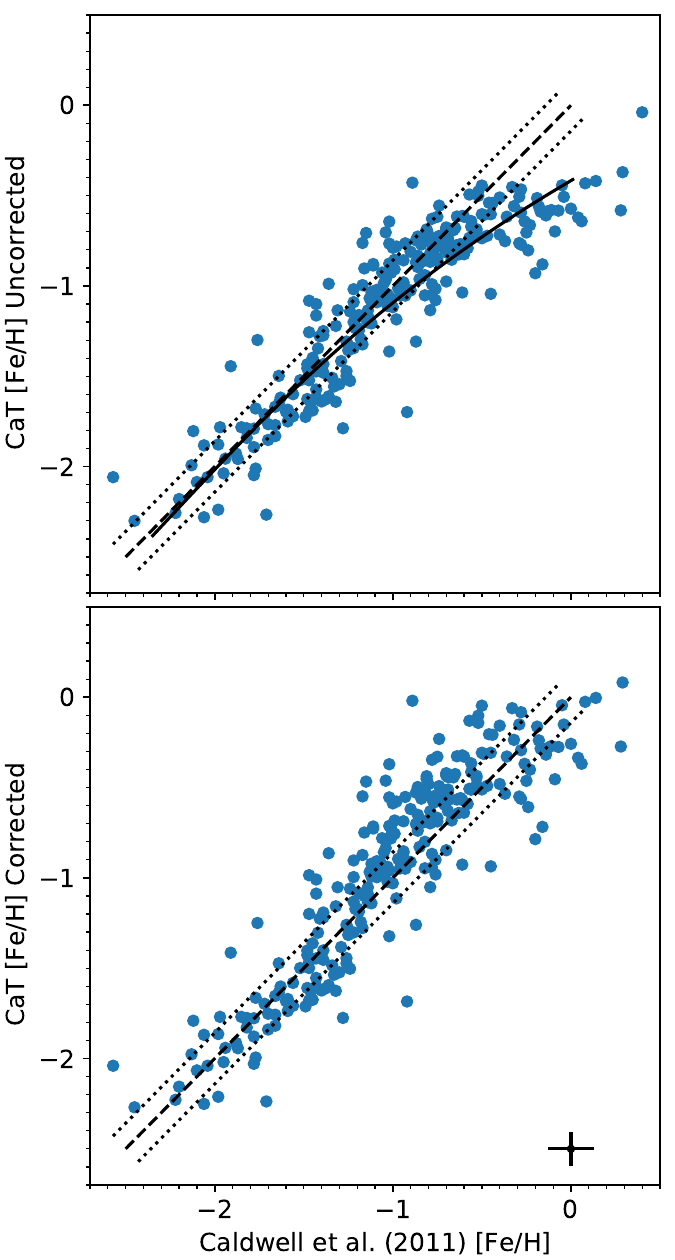}
    \caption{Comparison of the CaT [Fe/H] values measured from the \citet{2011AJ....141...61C} spectra and the [Fe/H] values measured by \citet{2011AJ....141...61C} using Lick Fe indices using the same spectra.
    The top panel shows the [Fe/H] calculated from the CaT values uncorrected for the effects of spectral resolution; the bottom panel shows the [Fe/H] calculated from the corrected CaT values.
    The solid line in the upper panel is the spectral resolution correction we apply to the CaT values.
    The median uncertainties are shown in the lower panel.
    The dashed line is the one-to-one and the dotted lines show $\pm 0.1$ dex.
    The metallicities measured using the two different methods agree within 0.25 dex.}
    \label{fig:caldwell_fe}
\end{figure}

Unfortunately, the measured strength of the CaT depends on the spectral resolution or velocity dispersion, with the CaT strength being systematically lower at lower spectral resolutions, especially at high metallicity \citep[e.g.][]{2001MNRAS.326..959C, 2019MNRAS.482.1275U}.
While the resolution of the Sakari et al. spectra are high enough not to be significantly affected by this effect, the \citet{2011AJ....141...61C} spectra are low enough resolution to be.
We used high S/N spectra of the CaT region of 27 GCs from the WAGGS survey \citet{2017MNRAS.468.3828U, 2019MNRAS.482.1275U} which span the range of MW GC metallicities ($-2.4 <$ [Fe/H] $< -0.1$) to derive an empirical correction for the effects of velocity dispersion on the measured strength of the CaT.
We first smoothed each of the WAGGS spectra to a range of velocity dispersions between 10 and 100 km s$^{-1}$ and measured the strength of the CaT on each smoothed spectra.
For each WAGGS GC we then fit a cubic spline between the measured velocity dispersion and the measured CaT strength.
We used these splines to estimate the CaT strength at a velocity dispersion of 72.3 km s$^{-1}$, the median of the velocity dispersions fitted to the \citet{2011AJ....141...61C} spectra, for each of the WAGGS GCs.
We then fit a cubic relationship between the CaT strength measured on the spectra broadened to 72.3 km s$^{-1}$ and the CaT strength measured on the original, unbroadened spectra.
This broadening is consistent with the line spread function of Hectospec ($\sigma = 78 \pm 4$ km s$^{-1}$) found by \citet{2013PASP..125.1362F} at 8600 \AA{} once the line spread function ($\sigma = 19$ km s$^{-1}$) of the DEIMOS template spectra is accounted for.
We used this relationship to correct each of the CaT strengths measured from the Caldwell spectra.
To estimate the systematic uncertainty introduced by this correction, we repeated the correction at 62.7 km s$^{-1}$ and 86.7 km s$^{-1}$, the 2.3 and 97.7 percentiles of the velocity dispersion distribution measured from the Caldwell et al. spectra.
We used these respectively to correct the upper and lower uncertainties on the measured CaT strength.
In terms of metallicity the strength of the correction varies from 0.09 dex at [Fe/H] $= -1$ to 0.42 dex at [Fe/H] $= 0$ as seen in Figure \ref{fig:caldwell_fe}.

We used the empirical correction of \citet[][their equation 1]{2019MNRAS.482.1275U} to translate our CaT values into metallicities.
The \citet{2019MNRAS.482.1275U} relation is based on MW and MW satellite GCs.
Since CaT is sensitive to the [$\alpha$/Fe] abundance ratio \citep{2012ApJ...759L..33B, 2016MNRAS.456..831S, 2019MNRAS.482.1275U}, if the [$\alpha$/Fe]-[Fe/H] relation of M31 GCs is different, our [Fe/H] measurements will be biased.
In Figure \ref{fig:caldwell_fe}, we show a comparison of our CaT based [Fe/H] with the [Fe/H] measured by \citet{2011AJ....141...61C} using Lick Fe indices and an empirical relationship before and after performing the spectral resolution correction.
The root mean squared (RMS) difference between the two metallicities is 0.25 dex, which is in agreement with the uncertainties.
While the correction for spectral resolution improves the agreement at the highest metallicities, small ($< 0.2$ dex) systematic differences remain, with the CaT metallicities being systematically higher near [Fe/H] $\sim -1.5$ and systematically lower at [Fe/H] $\sim -0.5$. 
We note that both the \citet{2011AJ....141...61C} metallicities and our CaT metallicities are based on simple empirical relationships between the strength of spectral indices and the metallicities of MW GCs.
Some level of systematic difference is unsurprising, given that the Fe Lick indices used by \citet{2011AJ....141...61C} and the CaT used in this work show difference dependencies on [$\alpha$/Fe] and the simple empirical relations used by \citet[bilinear]{2011AJ....141...61C} and \citet[linear]{2019MNRAS.482.1275U}.

\begin{figure}
    \centering
    \includegraphics[width=216pt]{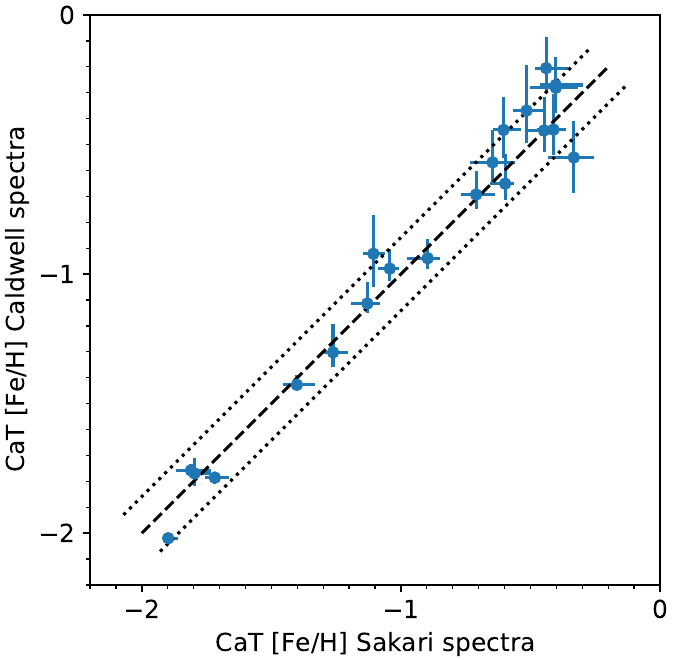}
    \caption{Comparison of the CaT [Fe/H] values measured in this work from the \citet{2016MNRAS.456..831S} spectra and the \citet{2011AJ....141...61C} spectra.
    The dashed line is the one-to-one and the dotted lines show $\pm 0.1$ dex.
    The metallicities measured using the two different sets of spectra agree within 0.1 dex.}
    \label{fig:caldwell_sakari_fe}
\end{figure}

\begin{figure}
    \centering
    \includegraphics[width=216pt]{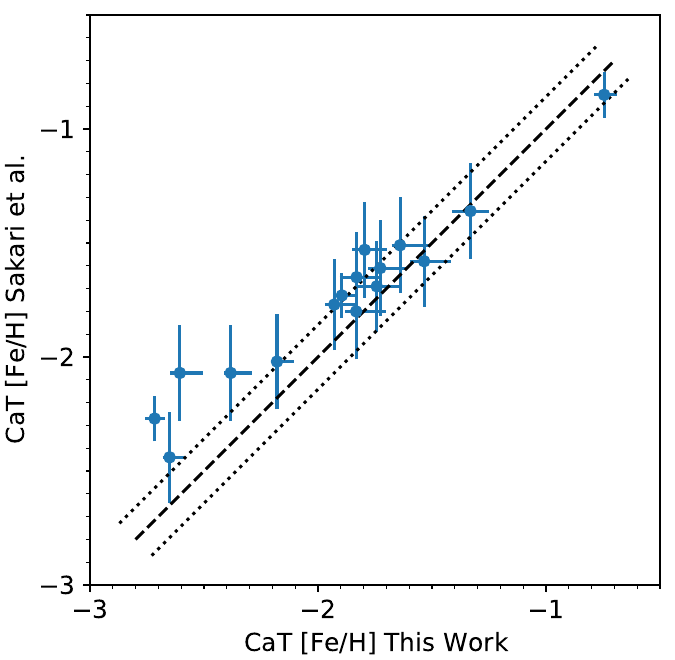}
    \caption{Comparison of the [Fe/H] values measured in this work and those measured by \citet{2022MNRAS.512.4819S} both using the strength of the CaT and the spectra from \citet{2022MNRAS.512.4819S}.
    The dashed line is the one-to-one and the dotted lines show $\pm 0.1$ dex.
    In general there is good agreement between the two studies except below [Fe/H] $= -2$ where the \citet{2022MNRAS.512.4819S} metallicities are systematically higher.}
    \label{fig:sakari_CaT_fe}
\end{figure}

\begin{figure*}
    \centering
    \includegraphics[width=398.4pt]{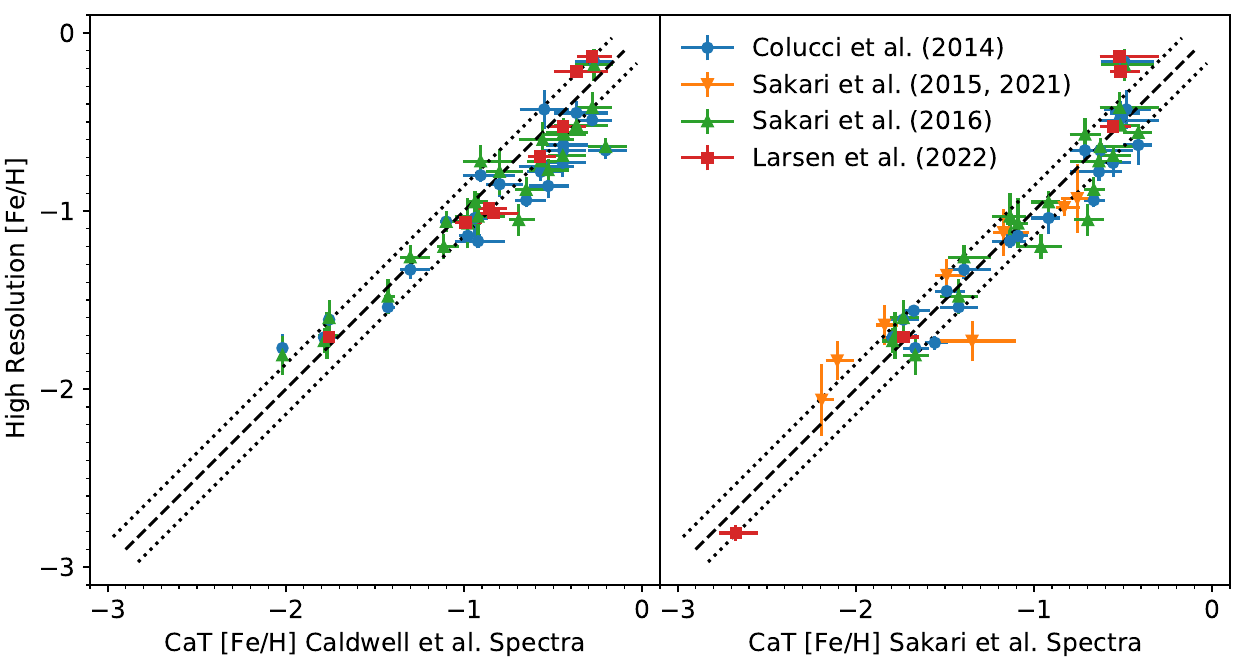}
    \caption{Comparison of the CaT [Fe/H] values with literature [Fe/H].
    The left panel shows a comparison with the Caldwell et al. spectra and the right with the Sakari et al. spectra.
    The dashed line is the one-to-one and the dotted lines show $\pm 0.1$ dex.
    There is generally good agreement between our metallicitiy measurements and those from the literature high resolution studies.}
    \label{fig:hires_fe}
\end{figure*}

In Figure \ref{fig:caldwell_sakari_fe} we show a comparison between the CaT [Fe/H] values measured from the \citet{2016MNRAS.456..831S} and \citet{2011AJ....141...61C} spectra.
The agreement between the two is excellent with a RMS difference of 0.11 dex in agreement with the statistical and expected systematic uncertainties.
In Figure \ref{fig:sakari_CaT_fe} we compare our CaT metallicities with those from \citet{2022MNRAS.512.4819S}.
Again we see good agreement at most metallicities although the \citet{2022MNRAS.512.4819S} metallicities are systematically higher at metallicities below [Fe/H] $= -2$.
Our measurement for EXT8 ([Fe/H] $=-2.71 \pm 0.07$) is closer to the value from high resolution spectroscopy (\citealt{2020Sci...370..970L}, [Fe/H] $= -2.91 \pm 0.04$) than the CaT based value from \citet{2022MNRAS.512.4819S} ([Fe/H] $= -2.27 \pm 0.10$).
In Figure \ref{fig:hires_fe} we compare our metallicities with those based on analysis of high resolution integrated light.
We see good agreement with the optical studies of \citet{2014ApJ...797..116C}, \citet{2015MNRAS.448.1314S} and \citet{2022A&A...660A..88L} as well as the near-infrared study of \citet{2016ApJ...829..116S} using either the Caldwell et al. or Sakari et al. spectra.

\section{Age measurements}
\label{sec:age}
To determine ages we used the Monte Carlo Markov Chain (MCMC) code first presented in \citet{2019MNRAS.490..491U} which we now name \textsc{mcmame} (Monte Carlo Metallicity, Age, Mass and Extiction).
In this analysis we sample age, metallicity, mass and reddening posteriors of a grid of stellar population models subject to the constraints provided by the photometry and optionally the CaT metallicity.
As in \citet{2019MNRAS.490..491U} we use \textsc{emcee} \citep{2013PASP..125..306F} with a parallel-tempered ensemble sampler \citep{2016MNRAS.455.1919V} with 8 temperatures, 64 walkers, 200 burn-in steps and 2560 production steps.
We use the same stellar populations models (\textsc{FSPS} \citealt{2009ApJ...699..486C, 2010ApJ...712..833C} calculated with the MIST isochrones \citealt{2016ApJ...823..102C} and the MILES spectral library \citealt{2006MNRAS.371..703S}), and mass priors (flat in log mass), but use slightly different age priors (flat between 1 Myr and 15 Gyr) and metallicity priors (flat between [Fe/H] $= -3.0$ and $+0.5$) than in \citet{2019MNRAS.490..491U}.
Unlike in \citet{2019MNRAS.490..491U} where the reddening vector from \citet{2011ApJ...737..103S} was assumed for all ages and metallicities, we used \textsc{FSPS} to calculate the age and metallicity dependent reddening vector using the \citet{1989ApJ...345..245C} MW extinction curve.
At low extinction values this assumption is valid but for larger values it breaks down since the effect of extinction depends on the shape of the spectral energy distribution within a filter passband, with a population with a bluer spectrum being more effected than a redder one.
For reddening we need to adopt different priors than in \citet{2019MNRAS.490..491U} given that M31 has substantial internal extinction, unlike the massive early-type galaxies studied in \citet{2019MNRAS.490..491U}.

As a first order approximation, we can think of most of the internal extinction in M31 coming from a thin disc of gas and dust.
The majority of old and intermediate age GCs should be either in front of or behind this disc as star clusters would not survive long orbiting in the same plane as massive molecular clouds \citep{2010ApJ...712L.184E, 2015MNRAS.454.1658K}.
Thus the prior on the extinction should be bimodal, with one peak corresponding to the foreground extinction and one peak corresponding to the extinction of the foreground plus the disc of M31.
In principle a GC should have equal probability of being in front or behind of the disc.
However, in a magnitude limited sample, GCs on the far side of the disc will be fainter and more likely to fall out of the selection.

The extinction can be modelled through various methods including from far infrared and sub-millimetre observations \citep[e.g.][]{1998ApJ...500..525S, 2019MNRAS.489.5436W} and by modelling the colours of stars \citep[e.g.][]{2015ApJ...814....3D}.
Unfortunately, the distribution of gas and dust is filamentary, with significant changes in the extinction value on comparable scales to the radius of a GC \citep[e.g.][]{2013MNRAS.435..263B, 2019ApJ...874...86S, 2020MNRAS.493.2688B}.
While for M31, extinction can be modelled on the scales of $\sim 30$ pc \citep[e.g.][]{2019MNRAS.489.5436W, 2015ApJ...814....3D}, for more distant galaxies it can only be modelled on larger scales.
To mimic the low spatial resolution available for more distant galaxies, we used the \citet{1998ApJ...500..525S} extinction maps which have a resolution 6.1 arcmin ($\sim 1.4$ kpc at the distance of M31).
We modelled the contribution of the disc of M31 to the extinction as a Gaussian with a mean provided by the value of the Schlegel map at that location and a standard deviation equal to 50 \% of the mean.
We used the median extinction from the Schlegel map of GCs outside of the disc of M31 (0.08 mag) as the mean value of the foreground contribution and assumed a dispersion of 0.03 mag on the foreground.
In our bimodal extinction prior we assigned equal weight to the foreground and M31 disc components.
In all cases we enforced a further constraint that the extinction must be positive.

\begin{figure*}
\includegraphics[width=504pt]{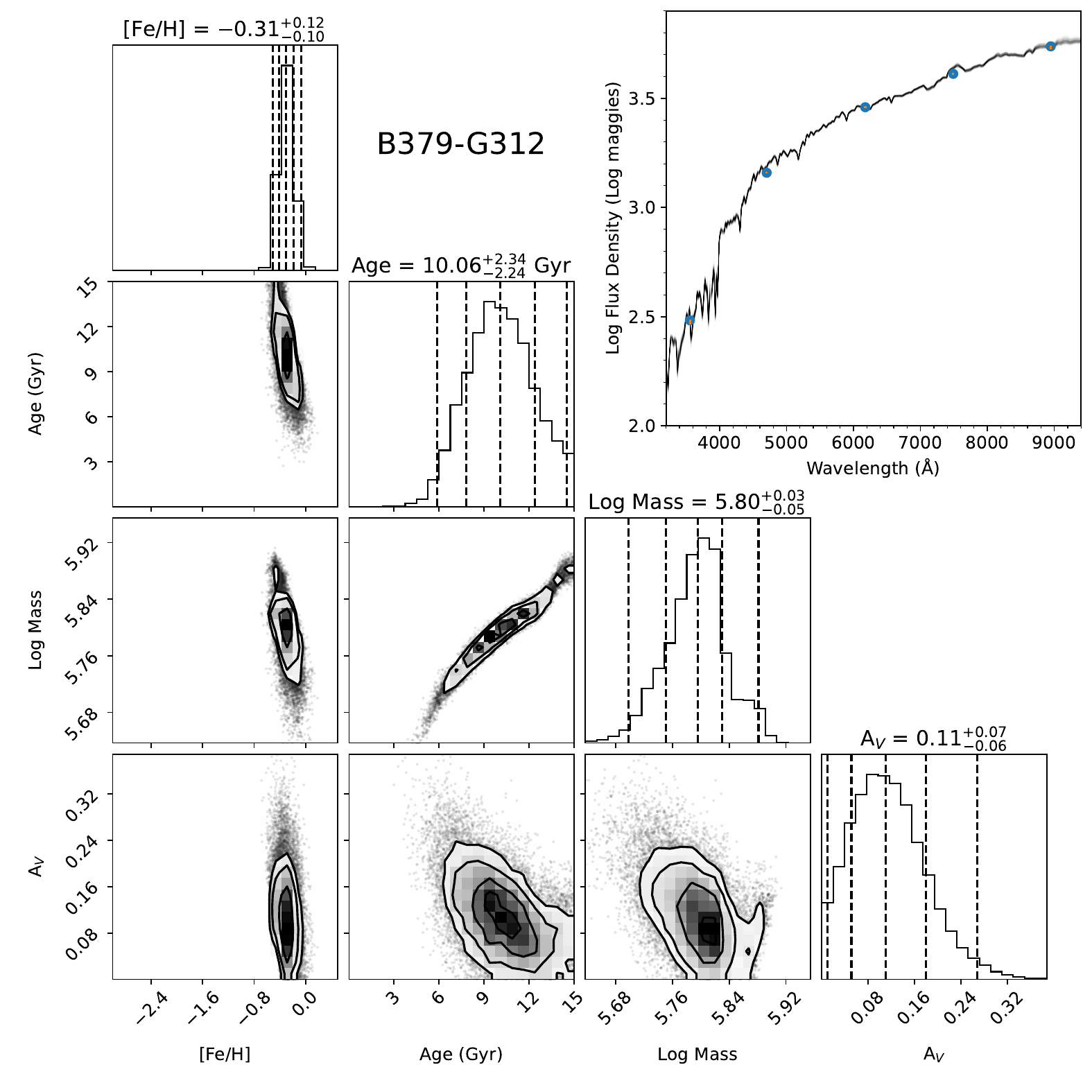}
\caption{Posterior distributions of metallicity, age, mass and reddenings for the M31 GC B379-G312 based on SDSS $ugriz$ photometry and a CaT based metallicity prior.
In the top right, we plot the spectral energy distributions calculated by \textsc{fsps} for 256 points drawn at random from the posterior distribution in grey, the median colours of the posterior distribution as blue circles and the observed photometry as orange error bars.
We note that B379-G312 presents the best case for fitting its photometry, as it is bright, metal rich and outside of the projected disc of M31.}
\label{fig:B379_corner}
\end{figure*}

\begin{figure*}
\includegraphics[width=504pt]{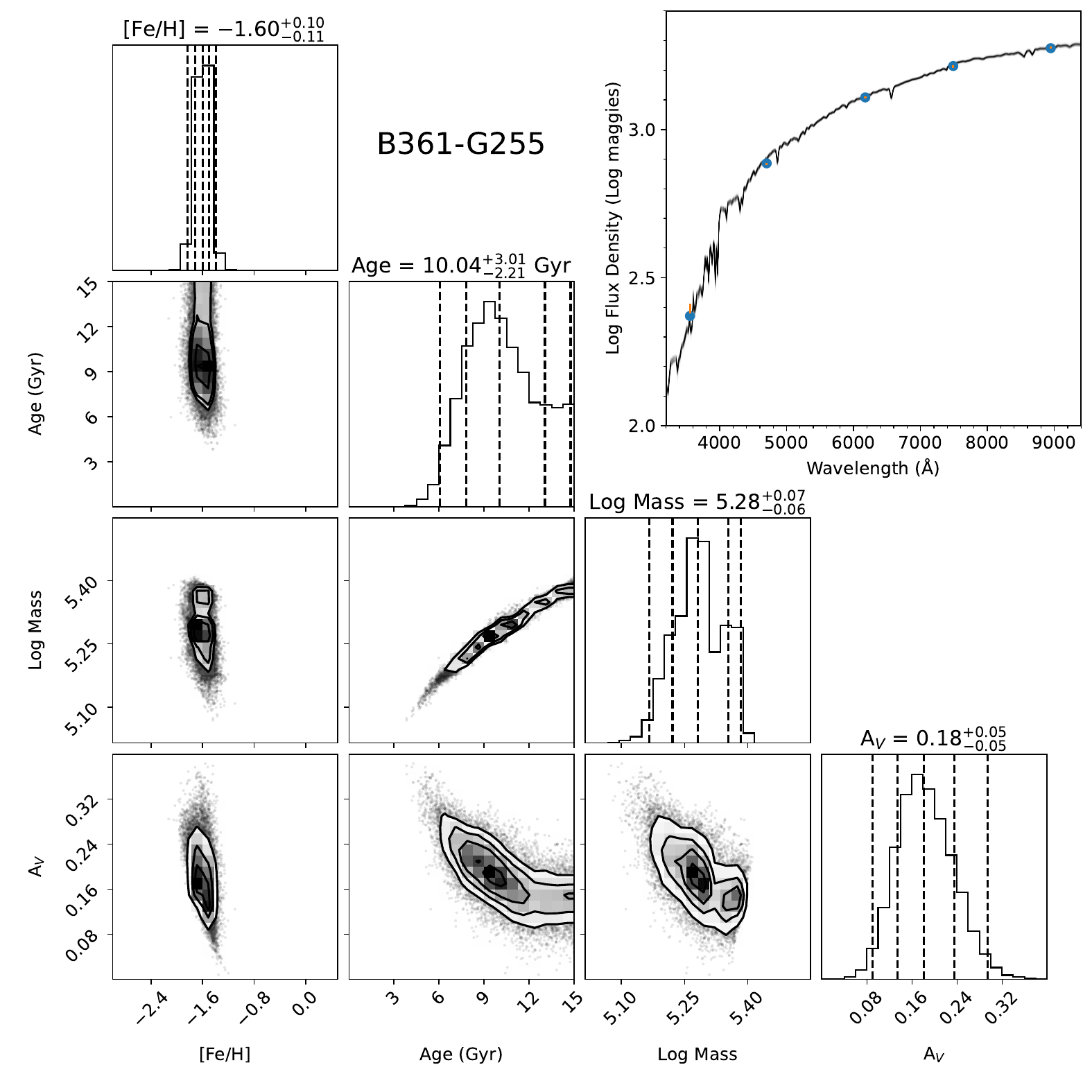}
\caption{Posterior distributions of metallicity, age, mass and reddenings for the M31 GC B361-G255 based on SDSS $ugriz$ photometry and a CaT based metallicity prior.
In the top right, we plot the spectral energy distributions calculated by \textsc{fsps} for 256 points drawn at random from the posterior distribution in grey, the median colours of the posterior distribution as blue circles and the observed photometry as orange error bars.
Unlike B379-G312 plotted in Figure \ref{fig:B379_corner}, B361-G255 presents a more typical case for fitting its photometry, due to its median brightness and metal poor nature.}
\label{fig:B361_corner}
\end{figure*}

An example of the posterior distribution for B379-G312 is plotted in Figure \ref{fig:B379_corner}.
This presents the best case for sampling the posterior distribution as B379-G312 is bright, metal rich and outside of the projected disc of M31.
A more typical case is show in Figure \ref{fig:B361_corner}, a low metallicity ([Fe/H] $= -1.60$) GC with an apparent $g$-band magnitude equal to the median of our sample ($g = 17.25$).
In Table \ref{tab:age_metal} we give our age, metallicity, mass and extinction posteriors, the full version of which is provided online as Supporting Information.
Our median metallicity uncertainty is 0.10 dex and our median age uncertainty is 2.4 Gyr.

\begin{table*}
    \centering
    \begin{tabular}{ccccccccccc}
        ID & RA & Dec & [Fe/H]$_{\text{CaT}}$ & $A_{V \text{prior}}$ & Mass & Age & [Fe/H] & $A_{V}$ & Photo & Spectra \\
         & [deg] & [deg] & [dex] & [mag] & [$\log$ M$_{\odot}$] & [Gyr] & [dex] & [mag] & & \\
        (1) & (2) & (3) & (4) & (5) & (6) & (7) & (8) & (9) & (10) & (11) \\ \hline 
        B001-G039 & 9.9626 & 40.9696 & $-0.53 \pm 0.15$ & $0.40 \pm 0.20$ & $5.73_{-0.06}^{+0.04}$ & $11.4_{-3.1}^{+2.5}$ & $-0.64_{-0.11}^{+0.09}$ & $0.89_{-0.09}^{+0.10}$ & P10 & C \\
        B002-G043 & 10.0107 & 41.1982 & $-2.16 \pm 0.09$ & $0.27 \pm 0.14$ & $5.10_{-0.09}^{+0.04}$ & $10.9_{-2.7}^{+2.7}$ & $-2.15_{-0.09}^{+0.09}$ & $0.29_{-0.04}^{+0.05}$ & P10 & C \\
        B003-G045 & 10.0392 & 41.1849 & $-1.47 \pm 0.11$ & $0.29 \pm 0.15$ & $4.93_{-0.06}^{+0.10}$ & $4.3_{-1.2}^{+3.1}$ & $-1.47_{-0.09}^{+0.10}$ & $0.58_{-0.13}^{+0.11}$ & P10 & C \\
        B004-G050 & 10.0747 & 41.3779 & $-0.57 \pm 0.16$ & $0.24 \pm 0.12$ & $5.46_{-0.03}^{+0.03}$ & $12.5_{-2.1}^{+1.7}$ & $-0.76_{-0.08}^{+0.07}$ & $0.25_{-0.07}^{+0.07}$ & P10 & C \\
        B005-G052 & 10.0847 & 40.7329 & $-0.68 \pm 0.16$ & $1.03 \pm 0.51$ & $6.06_{-0.08}^{+0.03}$ & $11.9_{-4.1}^{+2.2}$ & $-0.72_{-0.12}^{+0.11}$ & $0.44_{-0.11}^{+0.15}$ & P10 & C \\
        ... & ... & ... & ... & ... & ... & ... & ... & ... & ... & ... \\ \hline
    \end{tabular}
    \caption{Age and metallicity posteriors.
    Values provided median value of the posterior and the 68 \% confidence interval.
    Column (1): GC name.
    Column (2): Right ascension in deg.
    Column (3): Declination in deg.
    Column (4): Metallicity prior from the strength of the CaT in dex.
    Column (5): $V$-band extinction prior from \citet{1998ApJ...500..525S} in magnitudes.
    Column (6): Mass posterior in log solar masses.
    Column (7): Age posterior in Gyr.
    Column (8): Metallicity posterior in dex.
    Column (9): $V$-band extinction posterior in magnitudes.
    Column (10): Photometry source. P10: \citet{2010MNRAS.402..803P}, H14: \citet{2014MNRAS.442.2165H}, SDSS: SDSS DR17 \citep{2022ApJS..259...35A}.
    Column (11): Spectra source. C: \citet{2009AJ....137...94C}, S: \citet{2016MNRAS.456..831S, 2021MNRAS.502.5745S, 2022MNRAS.512.4819S}.
    The full version of this table is provided in machine readable form in the online Supporting Information. }
    \label{tab:age_metal}
\end{table*}

\begin{figure}
\includegraphics[width=240pt]{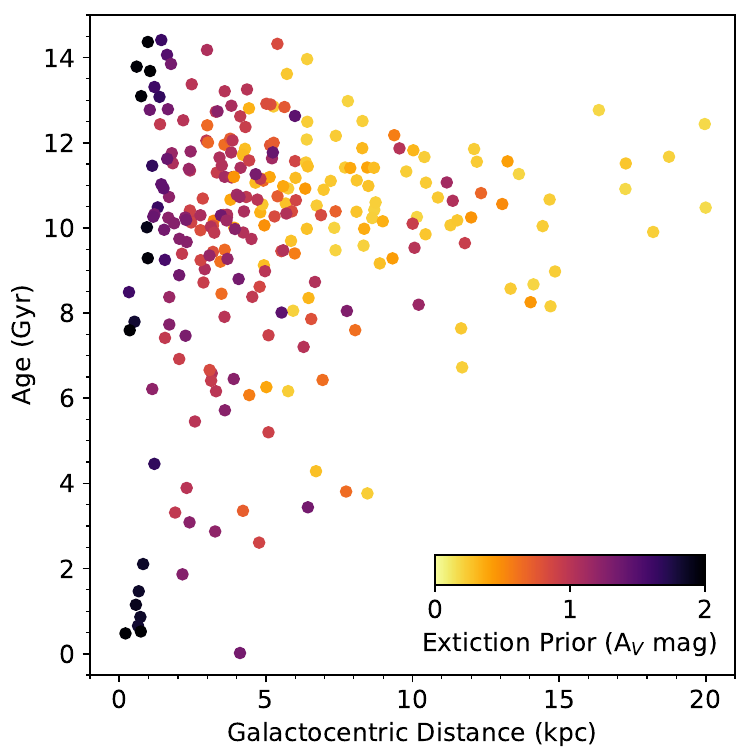}
\caption{Medians of the age posteriors versus galactocentric distance colour coded by the extinction prior from \citet{1998ApJ...500..525S} for GCs with spectra from \citet{2011AJ....141...61C}.
GCs with young age posteriors are found at small galactocentric distances and high possible extinction values.}
\label{fig:age_extiction}
\end{figure}

A number of GCs from the Caldwell sample show surprisingly young age posteriors, given that the Caldwell sample of spectra were selected to be older than a couple Gyr on the basis of their spectra between 3750 and 5400 \AA{} (see section 5 of \citealt{2009AJ....137...94C}).
The lack of young GC in both of the Caldwell and Sakari samples is supported by the lack of GC spectra with significant Paschen absorption which is seen in stellar populations younger than $\sim 1$ Gyr \citep[e.g.][]{2019MNRAS.482.1275U}.
In the case of some GCs such as M009, the age is poorly constrained, with the posterior distribution spanning a wide range of ages ($2.9_{-2.7}^{+8.8}$ Gyr in the case of M009 using the 16 to 84 \% percentile confidence interval.).
However, a number of GCs are inconsistent with being as old as 5 Gyr at a 95 \% level.
As can be seen in Figure \ref{fig:age_extiction}, these GCs with apparently young ages are found at small galactocentric radii and have large priors on their extinctions.
Thus these GCs are found in the crowded centre of M31, where the effects of the age-extinction degeneracy are strongest and both photometry and spectroscopy are most likely to suffer from systematic uncertainty.

\subsection{Comparison with previous measurements}

A wealth of studies have examined the ages of GCs in M31 using a range of techniques.
In the first class of studies, ages were constrained using resolved star colour-magnitude diagrams.
Only B379-G312 has HST imaging deep enough to reach the main sequence turn-off and measure the age directly.
For this GC (see Figure \ref{fig:B379_corner}) we find good agreement between our age measurement ($10.1_{-2.2}^{+2.3}$ Gyr) and the colour magnitude diagram based literature values ($10.0_{-1.0}^{+2.5}$ Gyr \citealt{2004ApJ...613L.125B}, $11.0 \pm 1.5$ \citealt{2010PASP..122.1164M} Gyr).
Shallower HST imaging can be used to place lower limits on the ages of M31 GCs from the non-detection of a main sequence or from the presence of blue horizontal branch stars \citep[e.g.][]{2011A&A...531A.155P}.
Using the MIST isochrones \citep{2016ApJ...823..102C} to estimate the brightness of the subgiant branch as a function of age and metallicity, we find no inconsistencies between the ages and metallicities we measure and the colour magnitude diagrams in \citet{2012A&A...546A..31P} for the 34 GCs in common.
Of the 12 GCs from \citet{2012A&A...546A..31P} with well measured HB properties, all GCs with blue HBs are older than 11.6 Gyr except for B350-G162 ($7.6_{-2.0}^{+4.0}$ Gyr) which is consistent with being old.
Published colour-magnitude diagrams for PA06, PA53, PA54 and PA56 \citep{2015MNRAS.448.1314S} and EXT8 \citep{2021A&A...651A.102L} all reveal blue horizontal branches, consistent with the old ages and low metallicities we measure.

Most studies of the ages of M31 GC have relied on integrated light.
Some studies have measured line indices and then fit stellar population models to the indices \citep[e.g.][]{2005A&A...434..909P, 2005AJ....129.1412B, 2011AJ....141...61C}.
Others have directly fit stellar population models to spectra \citep[e.g.][]{2013A&A...549A..60C, 2014ApJ...797..116C, 2015MNRAS.448.1314S, 2016AJ....152...45C}, to photometry \citep[e.g.][]{2009AJ....137.4884M, 2010ApJ...725..200F} or a combination there of \citep{2021A&A...645A.115W}.

\begin{figure*}
\includegraphics[width=504pt]{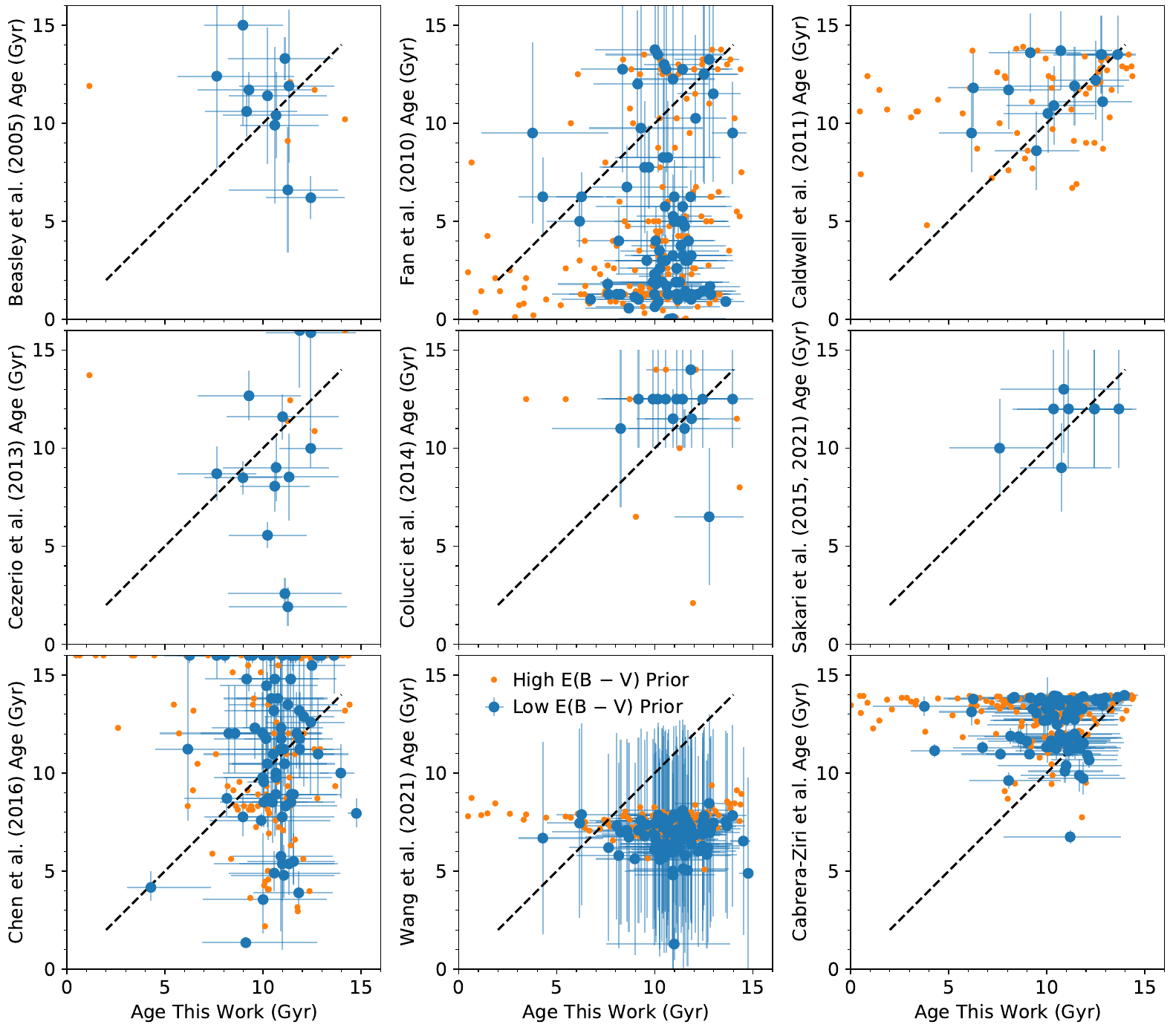}
\caption{Comparison of our MCMC ages with ages from \citet{2005AJ....129.1412B}, \citet{2010ApJ...725..200F}, \citet{2011AJ....141...61C}, \citet{2013A&A...549A..60C}, \citet{2014ApJ...797..116C}, \citet{2015MNRAS.448.1314S}, \citet{2016AJ....152...45C}, \citet{2021A&A...645A.115W} and Cabrera-Ziri et al. (in prep.).
For each study we show all GCs in common with high extinction priors ($E(B-V) > 0.15$ from \citealt{1998ApJ...500..525S}) as orange points and the GCs with low extinction priors ($E(B-V) < 0.15$) as blue circles with error bars.
When literature age measurements are larger than 16 Gyr we have plotted them at an age 16 Gyr.
The level of agreement varies between studies but in general we see good agreement.}
\label{fig:compare_literature_ages}
\end{figure*}

In Figure \ref{fig:compare_literature_ages} we show a comparison between our ages and those from literature.
In general we see better agreement for GCs with low extinction priors ($E(B-V) < 0.15$ from \citealt{1998ApJ...500..525S}) than for GCs with higher extinction priors which are projected onto the disc of M31.
This is unsurprising given the strong degeneracy between age and extinction for optical colours.
While this degeneracy is usually considered in studies of young star clusters, it is often ignored in studies of old globular clusters where the observed extinction is assumed only to be due to the Milky Way foreground.
While this assumption is acceptable in galaxy halos and in quiescent galaxies, it is not valid in actively star forming galaxies.

We compare our ages to ages measured by fitting stellar models to spectral indices \citep{2005AJ....129.1412B, 2011AJ....141...61C}, by directly fitting stellar population models to spectra \citep{2013A&A...549A..60C, 2014ApJ...797..116C, 2016AJ....152...45C} and by fitting stellar population models to photometry \citep{2010ApJ...725..200F}.
The level of agreement varies between studies, with the best agreement with \citet{2005AJ....129.1412B}, \citet{2011AJ....141...61C} and \citet{2014ApJ...797..116C} and the worst agreement with \citet{2010ApJ...725..200F} and \citet{2021A&A...645A.115W}.
\citet{2005AJ....129.1412B} used both the \citet{2003MNRAS.344.1000B} and the \citet{2003MNRAS.339..897T} models to convert Lick index measurements to ages and metallicities.
We find better agreement between our measurements and the \citet{2003MNRAS.339..897T} based values of \citet{2005AJ....129.1412B}; we display the \citet{2003MNRAS.339..897T} based ages in Figure \ref{fig:compare_literature_ages}.
\citet{2010ApJ...725..200F} fit $UBVRIJHK$ photometry with the \citet{2003MNRAS.344.1000B} models to derive ages and metallicities.
A comparison of the \citet{2010ApJ...725..200F} metallicity measurements and our own reveals that they typically over estimate metallicities which due to the age-metallicity degeneracy causes them to underestimate the ages.
\citet{2011AJ....141...61C} used EZAges \citep{2008ApJS..177..446G} to measure ages from Lick indices.
Due to the effects of the horizontal branch at lower metallicities, \citet{2011AJ....141...61C} only provide ages for GCs more metal rich than [Fe/H] $= -0.95$.
\citet{2013A&A...549A..60C} used full spectrum fitting and the \citet{2010MNRAS.404.1639V} stellar population models to measure ages.
\citet{2014ApJ...797..116C} and \citet{2015MNRAS.448.1314S} fitted their own stellar population models to high resolution spectra to measure ages, metallicities and abundances.
\citet{2016AJ....152...45C} measured ages using full spectral fitting both with the PEGASE–HR \citep{2004A&A...425..881L} and the MILES \citep{2010MNRAS.404.1639V} stellar population models.
The majority of PEGASE-HR based ages are either significantly younger or older than our age measurements while the MILES based ages show better agreement; we display the MILES based ages in Figure \ref{fig:compare_literature_ages}.
\citet{2021A&A...645A.115W} used a random forest classifier to derive ages and metallicities from Lick indices and $ugriz$ photometry.
They find ages of 6 to 8 Gyr for virtually all M31 GCs, inconsistent with our measurements and most other literature values.
Cabrera-Ziri et al. (in prep.) used \textsc{alf} \citep{Conroy12,Choi14,Conroy14,2018ApJ...854..139C,cabrera-ziri22} to fit spectra to measure ages, metallicities and abundances while accounting for the possibility of a hot horizontal branch.
While we see agreement for many GCs, most of Cabrera-Ziri et al. (in prep.) age measurements are a couple Gyr older than ours.

\begin{table}
    \centering
    \begin{tabular}{cccc}
Population & Age & [Fe/H] & N \\
 & [Gyr] & [dex] & \\
(1) & (2) & (3) & (4) \\ \hline
Combined & $10.7_{-0.2}^{+0.2}$ & $-1.00_{-0.01}^{+0.01}$ & 290 \\
Projected Disc & $10.3_{-0.3}^{+0.3}$ & $-0.84_{-0.02}^{+0.02}$ & 178 \\
Inner Halo & $10.9_{-0.4}^{+0.4}$ & $-1.22_{-0.03}^{+0.03}$ & 79 \\
Outer Halo & $12.5_{-0.5}^{+0.5}$ & $-1.77_{-0.03}^{+0.03}$ & 28 \\
NGC 205 & $10.8_{-1.1}^{+1.2}$ & $-1.43_{-0.06}^{+0.07}$ & 5 \\
Smooth Halo & $12.4_{-0.8}^{+0.7}$ & $-1.8_{-0.04}^{+0.04}$ & 16 \\
Ambiguous & $11.2_{-1.1}^{+1.2}$ & $-1.57_{-0.07}^{+0.06}$ & 5 \\
Substructure & $11.3_{-1.0}^{+1.1}$ & $-1.98_{-0.08}^{+0.09}$ & 7 \\
Ambiguous + Substructure & $11.3_{-0.7}^{+0.8}$ & $-1.63_{-0.07}^{+0.06}$ & 12 \\ \hline
Milky Way & $11.7_{-0.6}^{+0.5}$ & $-1.20_{-0.04}^{+0.04}$ & 65 \\
NGC 1407 & $11.9_{-0.3}^{+0.3}$ & $-0.60_{-0.02}^{+0.02}$ & 213 \\
NGC 3115 & $10.9_{-0.4}^{+0.4}$ & $-0.72_{-0.03}^{+0.03}$ & 116 \\
NGC 3377 & $8.0_{-0.4}^{+0.5}$ & $-0.80_{-0.06}^{+0.06}$ & 82 \\ \hline
    \end{tabular}
    \caption{Population age and metallicity medians from this study and from \citet{2019MNRAS.490..491U}.
    (1) Population name.
    (2) Median age in Gyr.
    (3) Median metallicity in dex.
    (4) Number of GCs.}
    \label{tab:age_metal_medians}
\end{table}
 
\subsection{Age-metallicity distributions}
\begin{figure*}
\includegraphics[width=504pt]{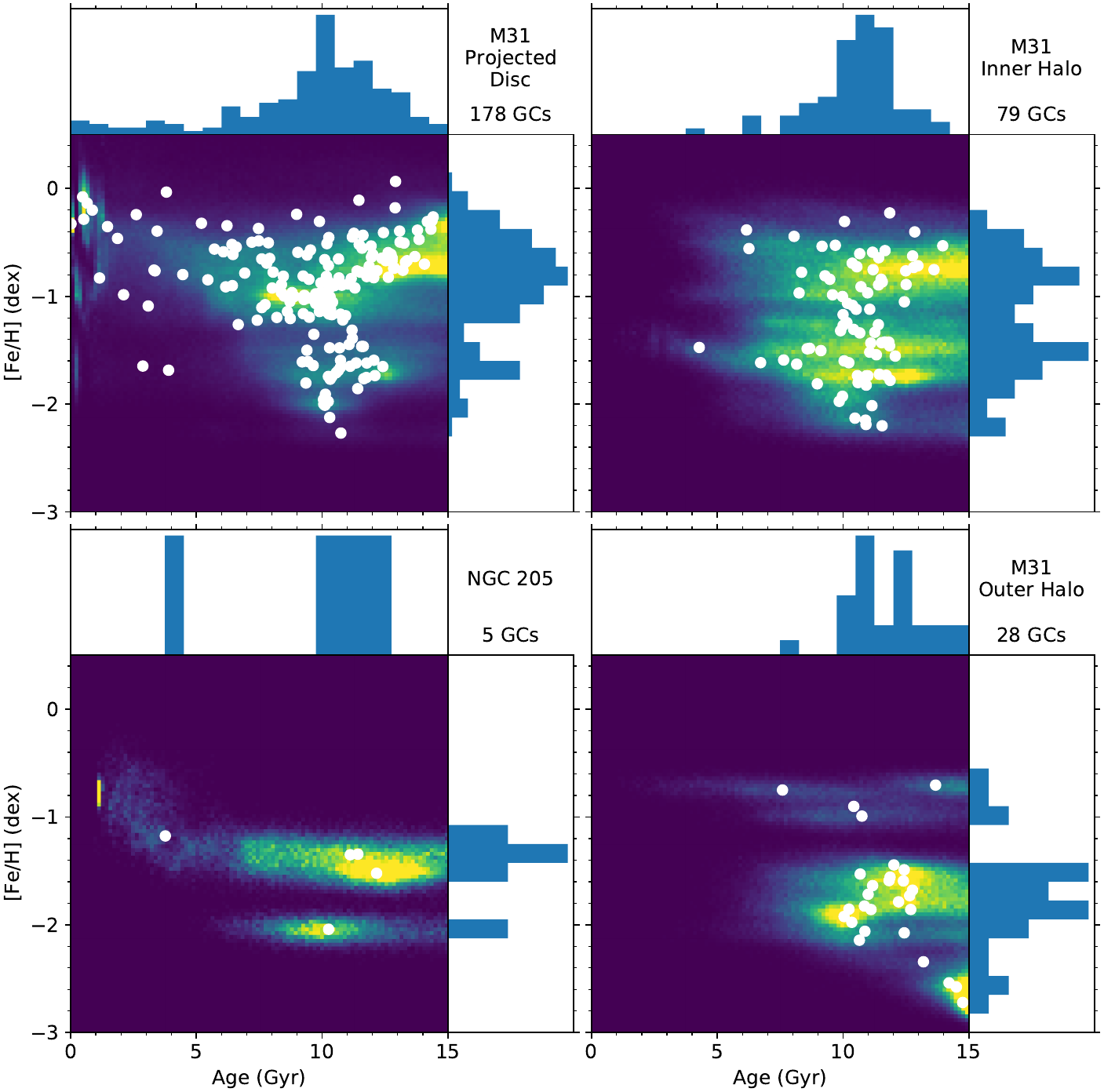}
\caption{Age-metallicity posterior distributions for M31 GCs based on photometry assuming a bimodal extinction prior.
The 2D histograms show the sum of the individual age-metallicity posterior distributions while the white points and the 1D histograms show the medians of the age and metallicity posteriors for each GC.
In the top left panel we show GCs projected on to the disc of M31, in the top right GCs in the inner halo of M31 (projected galactocentric distances $< 20$ kpc), in the bottom right  GCs in the outer halo of M31 (projected galactocentric distances $> 25$ kpc) and the bottom left GCs associated with the M31 satelite galaxy NGC 205.
For the outer halo subsample the ages and metallicities are on the Sakari et al. spectra and a mix of SDSS $ugriz$ and PAndAS $gi$ photometry; for the other subsamples they are based on the \citet{2011AJ....141...61C} spectra and the \citet{2010MNRAS.402..803P} SDSS $ugriz$ photometry.}
\label{fig:age_metal_subsamples}
\end{figure*}

In Figure \ref{fig:age_metal_subsamples} we show the age-metallicity distributions of GCs associated with different components of M31 while in Table \ref{tab:age_metal_medians} we give the median ages and metallicities for each of the components.
For GCs projected onto the disc of M31, which are based on the \citet{2011AJ....141...61C} spectra and \citet{2010MNRAS.402..803P} photometry and have relatively broad extinction priors, we see a wide range of ages and metallicities.
The wide range of ages is likely due to the age-extinction degeneracy.
For GCs in the inner halo of M31 - not projected on to the disc on M31 but within 20 kpc in projection - which are based on \citet{2011AJ....141...61C} spectra and \citet{2010MNRAS.402..803P} photometry and have relatively narrow extinction priors - we see generally old ages.
The outer halo, beyond a distance of 25 kpc in projection and based on the Sakari et al. spectra and a mix of PAndAS and SDSS photometry, also shows old ages.
Finally, the GCs associated with the dwarf elliptical NGC 205 show low metallicities.
Four of the NGC 205 GCs show old ages but the fifth, B330-G056, shows evidence for a younger age ($3.8_{-2.6}^{+3.9}$ Gyr).

In line with the previously observed \citep{2009A&A...508.1285G, 2011AJ....141...61C, 2022MNRAS.512.4819S} metallicity gradient for M31 GCs, we see the range of metallicities shift to lower metallicities and the fraction of metal poor GCs increase as we go out from the centre of M31.
Using the commonly used \textsc{GMM} code of \citet{2010ApJ...718.1266M}, the projected disc and inner halo populations as well as the combined sample all show significant ($p \leq 0.01$ as well a negative kurtosis and well-separated peaks) evidence for bimodality while the outer halo sample is too small to reliably use \textsc{GMM}.
\citet{2023MNRAS.519.5384P} discusses the emergence of GC metallicity bimodality in the context of GC formation and destruction, showing that in high mass galaxies like M31, the relative lack of GCs at intermediate metallicities is likely due to these GCs being preferentially disrupted rather than a deficit of cluster formation.

We also see a positive age gradient, with GCs projected onto the disc having younger ages and the GCs in the outer halo having older ages although it is unclear whether this is driven by the larger age uncertainties at smaller radii.
The less constraining the observations are, the more the age posterior will resemble the prior, with our uniform prior having a median age of 7.5 Gyr.
\citet{2016ApJ...824...42C} divided M31's GCs up by metallicity into three components.
The most metal poor ([Fe/H] $< -1.5$) component and the intermediate metallicity ( $-1.5 <$ [Fe/H] $< -0.4$) component show consistent ages ($10.9_{-0.3}^{+0.4}$ Gyr and $10.4_{-0.3}^{+0.3}$ Gyr respectively) while the most metal rich component ([Fe/H] $> -0.4$) shows a slightly younger median age ($8.3_{-1.6}^{+1.5}$ Gyr).
We note that all but three of the 26 metal rich GCs have large reddening priors so we can not rule out that the younger age median for the most metal rich GCs is due to their larger uncertainties.

In the projected disc sample we see a lack of GCs with metallicities [Fe/H] $\sim -1.2$ with median age posteriors older than 11 Gyr which are present at higher or lower metallicities.
It is unclear if this is driven by the larger age uncertainties in this metallicity range compared to lower metallicity GCs or a genuine lack of relatively old GCs in this metallicity range.
The most metal poor and metal rich GCs show the oldest ages, with some age posteriors having medians older than the age of the Universe.
Under the naive expectation that the metallicity of a galaxy increase over time, the most metal poor GCs should be the oldest.
However, it is also plausible that such metal poor GCs formed later, either in a less massive galaxy or less likely, in a galaxy that has recently accreated a large amount of near pristine gas.
We also note that the three most metal poor GCs in M31, all found in the outer halo, lie in a regime - more metal poor than any GCs in the Milky Way and similar to the faintest ultra faint galaxies \citep[e.g.][]{2019ARA&A..57..375S} - where stellar evolution models and stellar population models are not as well tested.
The old ages of the most metal rich GCs are more puzzling as the expectation is more metal rich GC should be younger than GCs that formed in the same progenitor.
Although many of the lower metallicity GCs likely formed later in lower mass progenitors, it is unlikely that no lower metallicity GCs formed in and survived from the progenitor(s) of the metal rich GCs.
We note than many of these old GCs have large extinction priors and lie at small galactocentric radii where the systematic and statistical uncertainty of both the spectra and photometry is higher (see Figure \ref{fig:age_extiction}).

If we had systematically underestimated the CaT metallicity priors we would systematically overestimate the ages.
Due to the low resolution of the spectra used, the CaT metallicity prior is less reliable at these high metallicities.
However, \citet{2019MNRAS.490..491U} also saw a trend of increasing age with metallicity for the most metal rich GCs in NGC 1407 and NGC 3115 (see Figure \ref{fig:all_galaxies} below) suggesting this issue is not due to the lower resolution spectra.
Since the CaT metallicity is sensitive to the $\alpha$-element abundance in general and the Ca abundance in particular, differences in the [Ca/Fe] ratio of high metallicity M31 GCs to the MW GCs used to calibrate the CaT-[Fe/H] relation could introduce a bias.
However, measurements of the $\alpha$-elements in M31 GCs \citep[e.g.][]{2014ApJ...797..116C, 2016ApJ...829..116S, 2022A&A...660A..88L} find similar ratios to MW GCs although only a handful of near solar metallicity GCs have been studied in M31.

Additionally, if the stellar population models we utilise predict too blue of a spectral energy distribution for a given age and metallicity, the resulting age constraint will be biased to older ages.
More subtly, the combination of incorrectly predicted colours and poorly constrained reddening could lead to an older age and lower reddening being a better fit than the correct age and reddening.
Again, differences in chemistry between the models and the data could produce colour differences.
The models we utilise have a scale solar abundance pattern while GCs generally have $\alpha$-element enhanced abundances and typical show internal spreads in the abundances of light elements, with some stars showing enhanced He, N and Na and depleted C and O \citep[see review by][]{2018ARA&A..56...83B}.
\citet{2009ApJ...694..902L, 2007MNRAS.382..498C} and \citet{2019ApJ...872..136C} all find an $\alpha$-element enhanced population is bluer in shorter wavelength colours such as $(u - g)$ and $(g - r)$ and redder in longer wavelength colours like $(r - i)$ and $(i - z)$.
Using models also based on the MIST isochrones and MILES stellar library \citet{2019ApJ...872..136C} find that $\alpha$-element enhanced models are a better fit to the colours of massive early-type galaxies than scaled solar models while noting the abundances of C and N also significantly affect the colours of an old, metal rich population.
\citet{2018MNRAS.478.2368C} modelled the effects of He and of enhanced N and Na with depleted C and O populations.
He enhanced populations are bluer in all optical colours (see also \citealt{2017ApJ...842...91C}) while N and Na enhanced, C and O depleted populations have the same $(u - g)$ colours as solar abundance models but bluer $(g - r)$, $(r - i)$ and $(i - z)$
We note the effects of abundance variations on colour are generally larger in older and higher metallicity populations since the effects are stronger in cooler stars \citep[e.g.][]{2009ApJ...694..902L}.
Further work is required to understand what effect different abundance ratios have on age estimates.

We note that there are larger systematic uncertainties in comparing ages at different metallicities than comparing ages at the same metallicity.
Given the larger systematic uncertainties at the extremes of the metallicity distribution and the larger statistical uncertainties for GCs in the projected disc, we do not attach any significance to the variation of the age of the oldest GCs with metallicity.

\begin{figure*}
\includegraphics[width=504pt]{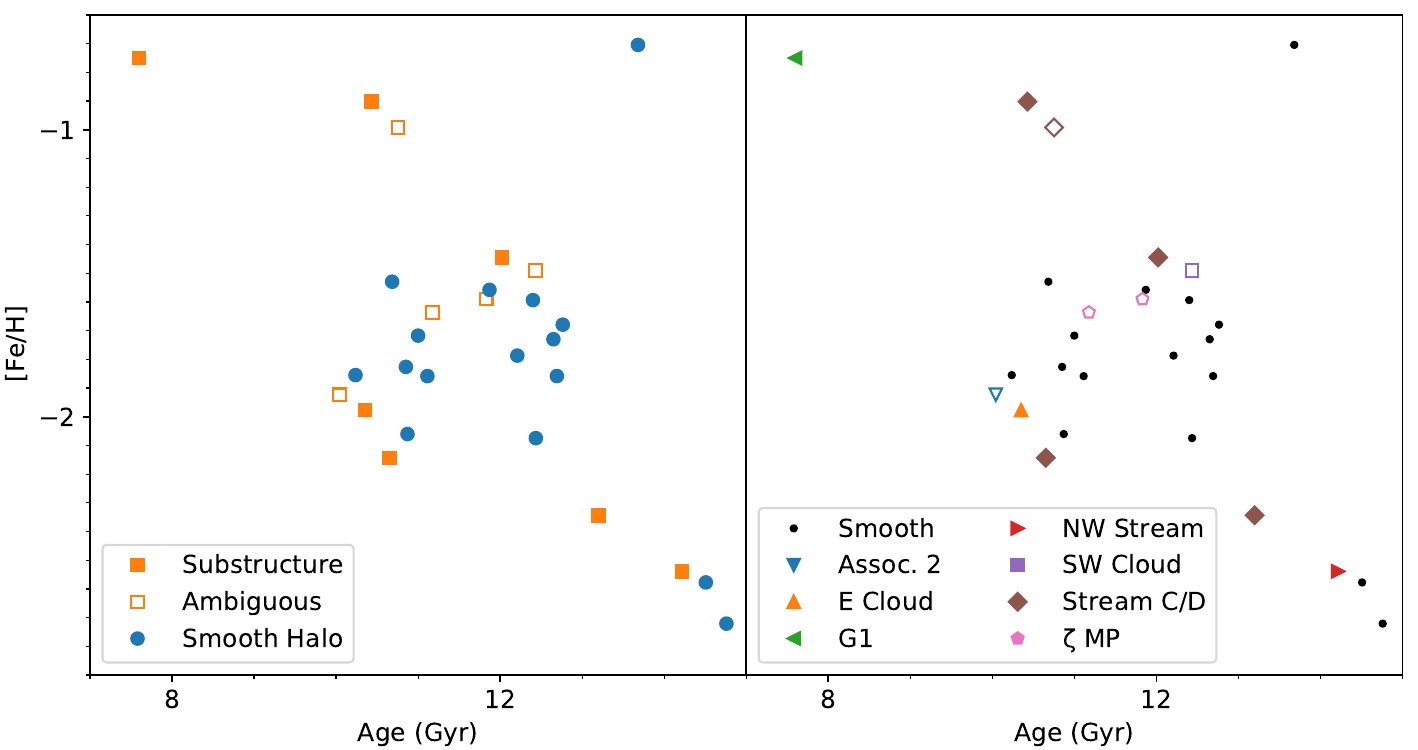}
\caption{Age-metallicity distribution of outer halo GCs.
In the left panel we show the GCs associated with the smooth halo by \citet{2019MNRAS.484.1756M} as blue circles and GCs associated with substructure as orange squares.
GCs classified as ambiguous by \citet{2019MNRAS.484.1756M} are denoted by open squares.
The youngest and most metal rich GCs are more likely to be associated with substructure than the smooth halo, inline with the predictions of \citet{2019MNRAS.482.2795H}.
In the right panel we show GCs associated with the smooth halo as black points while the GCs associated with substructure are shown as coloured polygons.
The colour and shape of these GCs corresponds to their classification by \citet{2019MNRAS.484.1756M} with GCs with ambiguous classifications having open symbols.
Unfortunately there are too few GCs associated with each substructure to robustly look for different age-metallicity relationships.}
\label{fig:halo_age_metal_by_stream}
\end{figure*}

For GCs in the outer halo, we used the association of \citet{2019MNRAS.484.1756M} between halo substructure and GCs to compare the GCs associated with the smooth halo with those associated with substructure.
As seen in the left panel of Figure \ref{fig:halo_age_metal_by_stream}, although there are old and extremely metal poor GCs associated with both the smooth halo and substructure, GCs associated with halo substructure extend to younger ages than those associated with the smooth halo.
The situation with metallicity is more complex.
Although the substructure and smooth halo GCs cover a similar range of metallicities and have similar medians, two of the seven GCs that are securely identified with substructure and one of the five ambiguous GCs are more metal rich than [Fe/H] $= -1.25$ while only one of the 16 smooth halo GCs is above this metallicity.
In Table \ref{tab:age_metal_medians} we give the medians of each of the populations.
The most metal rich smooth halo GC in our sample, PA-17, is older than the substructure GCs of similar metallicity.
That GCs with substructure and ambiguous classifications have higher metallicities than the those classified as part of the smooth halo has been previously noted by \citet{2022MNRAS.512.4819S}.
The younger ages and higher metallicities of some GCs associated with substructure supports the predictions of \citet{2019MNRAS.482.2795H} where the substructured halo has been more recently accreted than the smooth halo.
In the right panel of Figure \ref{fig:halo_age_metal_by_stream} we show the association of the substructure GCs with each halo feature; unfortunately there are too few GCs associated with each substructure to robustly study the age-metallicity relationship of each substructure.

\begin{figure*}
    \includegraphics[width=504pt]{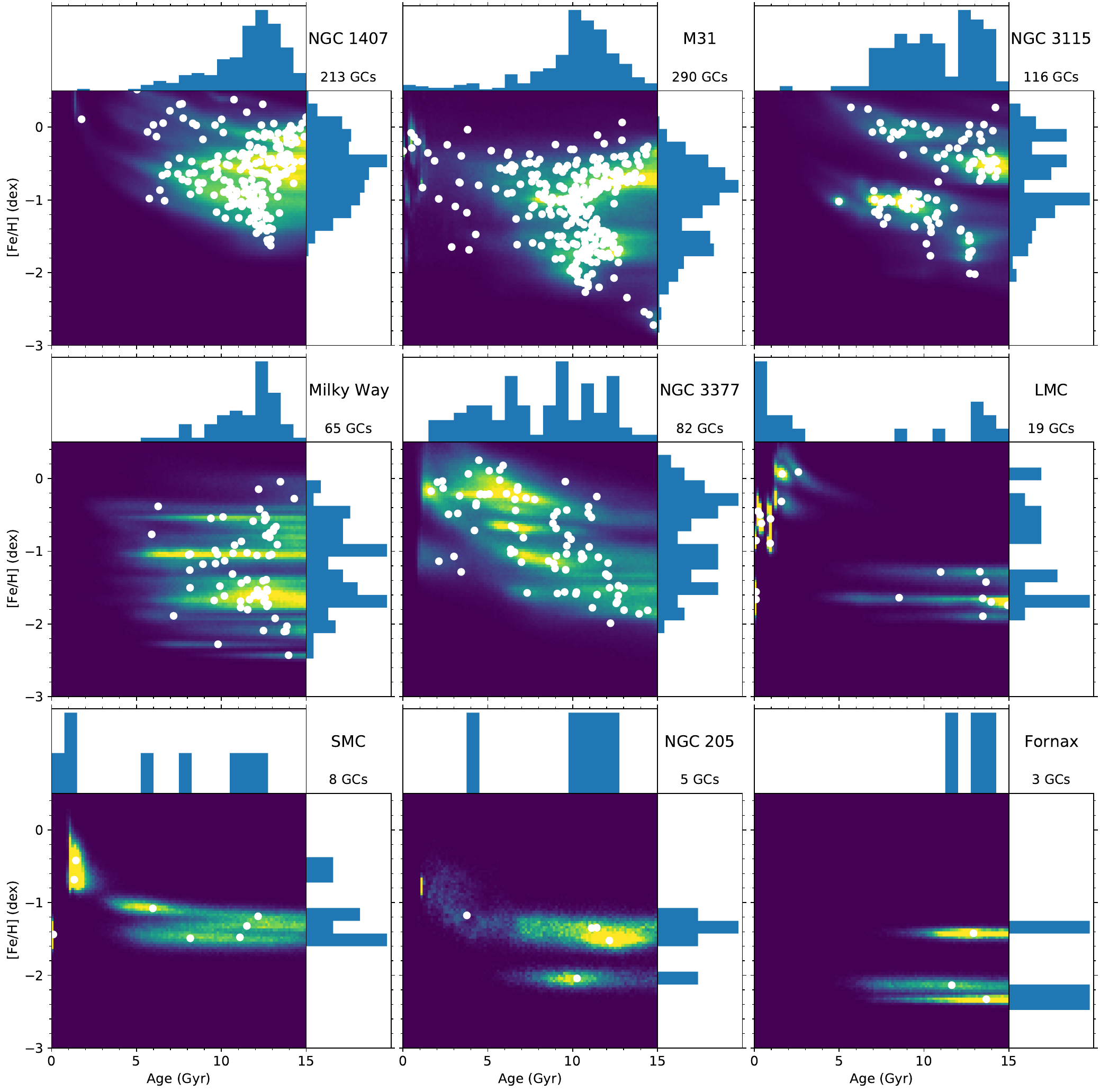}
    \caption{GC age-metallicity posterior distributions for the galaxies studied in this work (M31 and NGC 205) and the galaxies from \citet{2019MNRAS.490..491U}.
    The galaxies are ordered by stellar mass using masses from \citet{2017MNRAS.464.4611F} for the three SLUGGS galaxies, from \citet{2012ApJ...758...11C} for M31, from \citet{2016ARA&A..54..529B} for the MW and from \citet{2012AJ....144....4M} for the Local Group dwarf galaxies.
    As in Figure \ref{fig:age_metal_subsamples} the 2D histograms show the sum of the individual age-metallicity posterior distributions while the white points and the 1D histograms show the medians of the age and metallicity posteriors for each GC.}
    \label{fig:all_galaxies}
\end{figure*}

We can compare the age-metallicity distribution of M31 GCs with those of the galaxies studied by \citet{2019MNRAS.490..491U} in Figure \ref{fig:all_galaxies}.
We note that although the technique used by \citet{2019MNRAS.490..491U} is the same as in this paper, the GC selection and data quality differs between M31 and the \citet{2019MNRAS.490..491U} galaxies.
As noted by \citet{2019MNRAS.482.1275U}, the CaT is not a reliable metallicity indicator at ages younger than a couple Gyr when stages of stellar evolution other than the red giant branch or the red clump dominate the light in the CaT spectral region.
However, as noted by \citet{2019MNRAS.490..491U} due to weaker sensitivity of colour to metallicity at younger ages, our method still provides young ages for these star clusters.
For the giant galaxies, M31's GCs are slightly younger than those of NGC 1407 and the Milky Way, similar in age to those of NGC 3115 and older than those of NGC 3377 while being more metal poor than those of the three SLUGGS galaxies and more metal rich than those of the Milky Way.
For the dwarf galaxies, NGC 205's GCs show similar ages and metallicities to the old and intermediate age GCs in the similar mass SMC.

Although we defer a detailed and quantitative discussion of M31's formation and assembly history to future work, we can discuss M31 GC system in a qualitative manner similar to \citet{2019MNRAS.490..491U}.
That most of M31's GC are old suggests that M31 formed most of its stellar mass early, inline with the observed star formation history of M31's disc and spheroid \citep[e.g.][]{2008ApJ...685L.121B, 2015MNRAS.453L.113B, 2017ApJ...846..145W}.
The lack of clear branches in age-metallicity space does not have a strong implications on the assembly history of M31 given our large age uncertainties.
As can be seen in Figure \ref{fig:all_galaxies} with similar quality data we do not see the bifurcation of the MW's GCs into the older in situ branch and a younger accreted branch that more precise ages relieves \citep{2010MNRAS.404.1203F, 2013MNRAS.436..122L}.
Comparing M31 to the galaxies studied in \citet{2019MNRAS.490..491U} the similarities between the age and metallicity distributions of M31, NGC 1407 and the MW suggests some commonality in the assembly history of these galaxies compared to the extended formation of NGC 3377 and the bimodality of NGC 3115.

\section{Summary}
\label{sec:discussion}
Using the strength of the near infrared CaT spectral feature we have measured the metallicity of a large number of GCs in M31.
In line with previous work \citep[e.g.][]{2016MNRAS.456..831S, 2019MNRAS.482.1275U}, we find that the CaT is a reliable measure of metallicity although its reliability declines at high metallicity and at the lower spectral resolution of the \citet{2009AJ....137...94C} Hectospec spectra. 
We used these metallicities as priors when fitting optical photometry with stellar population models to measure the ages, metallicities and masses of the GCs.
Due to the strong degeneracy between extinction and age, the ages of GCs projected on to the disc of M31 are less reliable than those in the halo of M31.
We find good agreement between our age measurements with most but not all literature studies.
Most of the GCs projected on to the disc and virtually all halo GCs are old (consistent with ages $> 10.5$ Gyr, formation redshifts $z > 2$).
The distribution of GC ages and metallicities is similar but not identical to galaxies with a similar stellar mass.
The old ages of the majority of M31's GCs suggest that M31 formed much of its stellar mass early, inline with observations of the field star formation history.
In the outer halo we find the most metal rich and youngest GCs are more likely to be associated with substructure rather than the smooth halo, in line with the predictions of \citet{2019MNRAS.482.2795H}.

The technique used in this paper of combining photometry with a prior from spectroscopy is likely not the best way of measuring the ages of M31 GCs where high S/N optical spectra are available (see Cabrera-Ziri et al. in prep.).
Instead it serves as a test of a technique suitable for more distant galaxies.
Upcoming highly multiplexed red and near infrared spectrographs, such as MOONS \citep{2014SPIE.9147E..0NC, 2018SPIE10702E..1GT} on the Very Large Telescope, will allow large numbers of extragalactic GCs to be observed efficiently.
Like other techniques that rely on photometry, our method suffers from the age-extinction degeneracy and will perform worse when there is not a strong prior on extinction, such as in or near areas of active star formation.
In addition, spectra, even when only available in the red or near infrared, do provide some age information.
This technique should be refined by either fitting the spectra with stellar population models and using that age-metallicity posterior as a prior when fitting the photometry or by simultaneously fitting the photometry and spectroscopy with a stellar population model.

\section*{Acknowledgements}
We thank the anonymous referee for their helpful comments which improved the manuscript.
We also wish to thank Joel Pfeffer for useful discussions.
CU acknowledges the support of the Swedish Research Council, Vetenskapsr{\aa}det.
This study was supported by the Klaus Tschira Foundation.
We wish to thank Charli Sakari for making her spectra available.

This work made use of \textsc{numpy} \citep{numpy}, \textsc{scipy} \citep{scipy}, \textsc{matplotlib} \citep{matplotlib} and \textsc{corner} \citep{corner} as well as \textsc{astropy}, a community-developed core Python package for astronomy \citep{2013A&A...558A..33A}.
The CaT fitting code of \citet{2019MNRAS.482.1275U} relies on \textsc{pPXF} \citep{2004PASP..116..138C, 2017MNRAS.466..798C}.

Funding for the Sloan Digital Sky Survey IV has been provided by the Alfred P. Sloan Foundation, the U.S. Department of Energy Office of Science, and the Participating Institutions. 
SDSS-IV acknowledges support and resources from the Center for High Performance Computing at the University of Utah.
The SDSS 
website is www.sdss.org.
SDSS-IV is managed by the Astrophysical Research Consortium for the Participating Institutions of the SDSS Collaboration including the Brazilian Participation Group, the Carnegie Institution for Science, Carnegie Mellon University, Center for Astrophysics | Harvard \& Smithsonian, the Chilean Participation Group, the French Participation Group, Instituto de Astrof\'isica de Canarias, The Johns Hopkins University, Kavli Institute for the Physics and Mathematics of the Universe (IPMU) / University of Tokyo, the Korean Participation Group, Lawrence Berkeley National Laboratory, Leibniz Institut f\"ur Astrophysik Potsdam (AIP),  Max-Planck-Institut f\"ur Astronomie (MPIA Heidelberg), Max-Planck-Institut f\"ur Astrophysik (MPA Garching), Max-Planck-Institut f\"ur Extraterrestrische Physik (MPE), National Astronomical Observatories of China, New Mexico State University, New York University, University of Notre Dame, Observat\'ario Nacional / MCTI, The Ohio State University, Pennsylvania State University, Shanghai Astronomical Observatory, United Kingdom Participation Group, Universidad Nacional Aut\'onoma de M\'exico, University of Arizona, University of Colorado Boulder, University of Oxford, University of Portsmouth, University of Utah, University of Virginia, University of Washington, University of Wisconsin, Vanderbilt University, and Yale University.

\section*{Data Availability}
The \citet{2009AJ....137...94C} spectra are available from \url{https://oirsa.cfa.harvard.edu/signature_program/}.
The \citet{2016MNRAS.456..831S}, \citet{2021MNRAS.502.5745S} and \citet{2022MNRAS.512.4819S} spectra are available upon request from Charli Sakari.
The \citet{2010MNRAS.402..803P} and \citet{2014MNRAS.442.2165H} photometry was taken from those papers; the remaining SDSS DR17 \citep{2022ApJS..259...35A} photometry is available from \url{http://skyserver.sdss.org/dr17}.

\bibliographystyle{mnras}
\bibliography{bib}{}

\end{document}